\documentstyle[pre,aps,epsfig,twocolumn,amsmath]{revtex}        

\renewcommand{\epsilon}{\varepsilon}

\begin{document}

\draft
\twocolumn[\hsize\textwidth\columnwidth\hsize\csname @twocolumnfalse\endcsname
\title{Microscopic realizations of the Trap Model}

\author{I. Junier and J. Kurchan}
\address{\it  PMMH UMR 7636 CNRS-ESPCI,\\
 Ecole Sup{\'e}rieure de Physique et Chimie Industrielles,
\\
10, rue Vauquelin, 75231 Paris CEDEX 05,  France}

\date{\today}

\maketitle

\begin{abstract}

Monte Carlo
 optimizations of Number Partitioning and of Diophantine approximations
are  microscopic realisations  of `Trap Model' dynamics. This offers a fresh look at the physics
behind this model, and points at other situations in which it may apply.
  Our results strongly suggest that in any such realisation of the Trap Model, the 
response and correlation functions of smooth observables obey the fluctuation-dissipation
theorem even in the aging regime.
Our discussion for the Number Partitioning problem may be relevant for the 
class of optimization problems whose  cost function does not scale linearly with the size,
and are thus awkward from the statistical mechanic point of view.

\end{abstract}

\vskip2pc]

\narrowtext

\section{Introduction}
During the last half century different models  have been proposed
to capture the dynamical properties of glasses. 
Among the phenomenological approaches,
 an illuminating and extensively studied minimal one  is the `Trap Model' proposed by
Bouchaud \cite{Bouchaud}. It is based on the following
picture: a particle evolves in a landscape energy
that resembles a golf course with  holes
whose depths $h$ are exponentially distributed following $\sim e^{-\beta_c h}$:
\vspace{0.3cm}
\begin{figure}
\begin{center}
\psfig{file=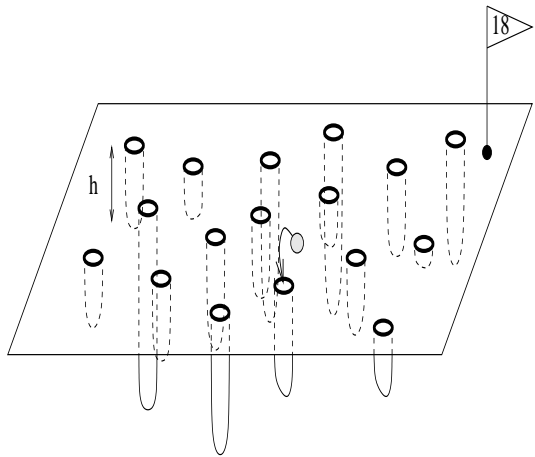,width=7cm,height=6cm}
\caption{Schematic view  of the phase space}
\label{fig1}
\end{center}
\end{figure}
At each microscopical time, the particle emerges from 
a hole to a {\em horizon level}  with a probability $\sim e^{-\beta h}$. Once escaped,
it immediately falls in a new  hole, which is in many applications randomly chosen
(i.e.  Fig. \ref{fig1} is 
infinite dimensional).  
By construction,
once the particle escapes from a hole, it totally loses memory and
the system can be considered as reinitialized.
The combination of exponential distribution of depths with an exponential
probability of emerging yields  trapping times following a distribution
whose mean diverges 
 if $ T < \beta_c^{-1}$.
Glassy behavior results from the possibility of falling in holes
with arbitrary long trapping times. We have the apparent paradox that if we consider
any two times such that the system has just emerged from a trap, the dynamics is
(statistically) reversible in time: irreversibility arises only because at the end
of a fixed  interval  the system is most probably in a long-lived trap.

These elements, existence of a horizon level
up
 to which the system has to reemerge each time~\footnote{ The reader familiar with 
mean-field glass models will wonder if this coincides with the `threshold' level arising there.
In fact, it does not: dynamics just below the threshold does not involve either 
reinitialization or reemerging, except possibly at very late times},
time reversibility, and the exponential density of states
can be taken as the defining features of the model. 
Even if such particle behaviors  
have not been observed, this phenomenological
approach gives many glassy features we can find in
both structural and spin glasses.
On the other hand,  only recently
trap model dynamics was shown to result for a microscopic model (the Random Energy Model)
 endowed with a microscopic  
stochastic dynamics \cite{Benarous} --- somewhat surprisingly
given that the original motivation
 for the model was the phase-space structure of mean-field glass models.

Having microscopic realizations of the trap model is interesting, because it highlights
physical situations when  it may be relevant. There is a further reason: 
 when it comes to studying the response of the dynamics to perturbations,
the trap model allows a wide range of possibilities \cite{BD,Sollich,Ritort}: the reason
for this  is that, although it is quite clear how  traps are affected by a perturbation,
there is no unique prescription for how a perturbing field should affect the `horizon',
and through it the transition probabilities. This is an 
important issue, since response functions (especially the Fluctuation-dissipation relation)
have turned
out to be amongst the most elloquent observables of glasses \cite{review}. 

In this paper we study two examples yielding trap model dynamics: the Number Partitioning
Problem --- itself a form of packing problem, and 
Diophantine Approximations. 
The  Number Partitioning
Problem can be stated as follows: given $N$ random numbers
drawn from a uniform distribution 
$[0,1]$, divide them into two sets such that their sums are as close as possible.
This is a caricature packing problem because if we consider $N$ coins with independent
random thicknesses
, this corresponds to finding the minimal
box height where they will fit in two piles. 
The Diophantine Approximations we shall consider is of the following kind:
find integer numbers $n$ and $m$ such that  $n{\sqrt 2}+m {\sqrt 3}$ is as close as possible
to an even integer.
 
The paper is organised as follows: in section II we study the number partitioning problem
from the dynamical point of view. In section III we briefly 
describe how the problem of improving
Diophantine approximations yields the same kind of dynamics. In the final section we
use these results to discuss which features seem  necessary 
for a model to have Trap Model-like dynamics, and we point out that when this happens, the
 fluctuation-dissipation theorem is
obeyed, at least for smooth observables.

\section{Number partitioning optimization and the trap model}

From an optimization point of view, the unconstrained Number Partitioning
Problem  is the following: 
given a set of real numbers ${a_1,a_2,...,a_N}$, each one belonging to $[0,1]$, find a
partition composed of two subsets such that the difference between
the sum of numbers over the two subsets is
as small as possible. 
It can be written as 
an infinite range Ising spin glass with Mattis-like antiferromagnetic couplings   
$J_{ij} = - a_i a_j $
\begin{equation}
H_{Mattis}=E_m^2=\sum_{i,j=1}^N s_i a_i a_j s_j
\label{Emattis}
\end{equation}
or, alternatively
\begin{equation}
E_m=\left| \sum_{i=1}^N a_i s_i \right|
\label{Em}
\end{equation} 
The  ground state (for typical $a_j$'s) of
(\ref{Em}) 
scales like $\langle {E_m}_o \rangle \propto \sqrt N 2^{-N}$ \cite{Mertens}. 
This means that  in the form (\ref{Emattis})
of the
Hamiltonian, the system is such that the interesting behaviour 
 occurs at exponentially 
small temperature in terms of  the size of the system.
In order to avoid this problem, in this paper we shall use a modified  
Hamiltonian, whose ground state energy is extensive in the thermodynamic limit: 
\begin{equation}
E= T_c \log (E_m) =  T_c \log \left| \sum_{i=1}^N a_i s_i \right|
\label{Ham}
\end{equation}
$T_c$ is an energy scale and the ground state energy becomes $\sim -T_c \ln 2 \;  N$.

\subsection{Equilibrium analysis}

Non-extensive optimization problems, for which the optimum does not
scale linearly with the system size, are not easily cast
in the formalism of  statistical mechanics.  For the number partitioning problem,
Mertens 
\cite{Mertens} proposed a clever bypass: instead of
working with a partition function defined on the basis of (\ref{Emattis}) or (\ref{Em})
and following the usual steps of computing the average free energy, he used statistical
means to study directly the independence of energy levels. 
Thus, he argued  that in the case of unconstrained number Partitioning problem,
 for large values of N
 energies ${E_m}$ are random 
variables distributed according to:
\begin{equation}
P(E_m)=\frac{2}{\sqrt{2\pi \sigma^2 N}}\exp 
\left(-\frac{E_m^2}{2\sigma^2 N} \right). \Theta(E_m)
\end{equation}
where $\Theta(E_m)$ is the step-function and $\sigma ^2 = \langle a^2 \rangle$,
and that for  ${E_m}<O (\sqrt N)$ they are   independent.
This result was later shown rigorously for the lowest $k$ energies \cite{microsoft}, and
studied in great detail in all regions in \cite{Silvio}.

\begin{figure}
\begin{center}
\psfig{file=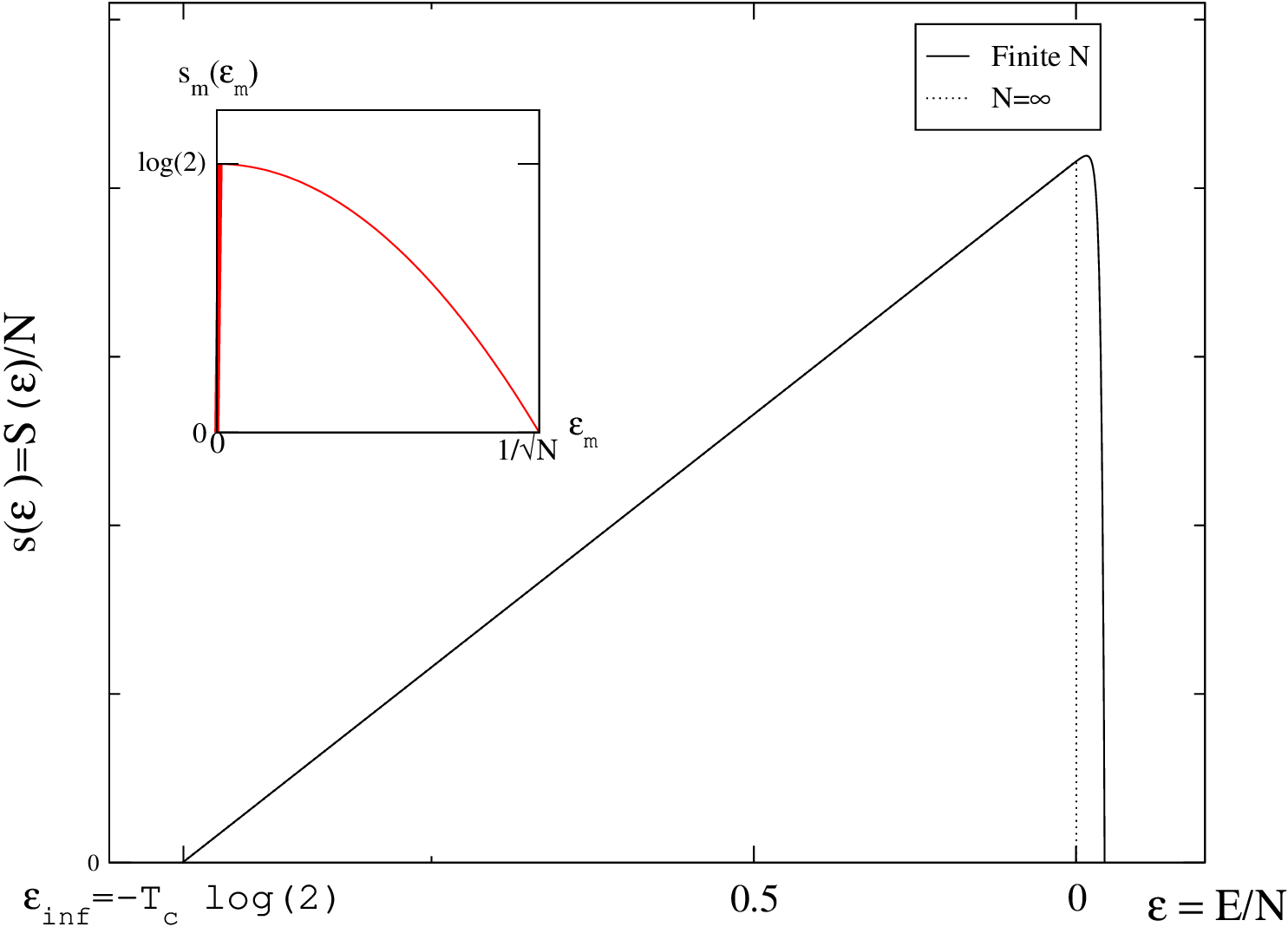,width=6cm,height=8cm,angle=-90}
\caption{
Entropy per spin $s=\frac{S}{N}$ vs energy per spin $\epsilon=\frac{E}{N}$.
Inset: corresponding original NPP}
\label{fig2}
\end{center}
\end{figure}

 From this it follows, applying the   transformation $E = T_c \log \vert E_m \vert$, 
that energies $E < O(\log N)$ are also independent random variables with a probability
distribution given by:
\begin{equation}
 P(E)= K(N)\exp \left(\beta_c E  - \frac{1}{2 \sigma^2 N} \exp \left(2 \beta_c E\right)\right)
\label{energie}
\end{equation}
with $K(N)=\frac{2 \beta_c}{ \sqrt{2 \pi \sigma^2 N}}$.
For large $N$, the density of levels is shown in Fig. \ref{fig2}.

\vspace{0.3cm}

The thermodynamics of the Number Partitioning Problem written in terms
of the new energy $E$ with  assumptions
above can be obtained  following 
 Derrida's microcanonical derivation for the Random Energy Model \cite{Derrida}. 
One first notices that there are two energy regions
\begin{itemize}
\item{If $E>E_{inf}=-N T_c \log 2$, the density of levels is much larger than 1.}
\item{If $E<E_{inf}=-N T_c \log 2$, the density of levels is much smaller than 1.}
\end{itemize}
For $E<E_{inf}$ the entropy vanishes, and for  $E \geq E_{inf}$, the entropy reads,
as $N \rightarrow \infty$:
\begin{equation}
\frac{S(E)}{N}=\log2+\frac{\beta_c E}{ N} \Theta(-E)
\end{equation}
Using the relation $\frac{dS}{dE}=\frac{1}{T}$, one finds that
$\frac{E}{N}$=0 when $T>T_c$, and that  for $T<T_c$, the energy sticks at 
$E_{inf}=-T_c N \log 2$. Finally, the free energy reads:
 \begin{eqnarray}
\left\{ \begin{array}{c}
\frac{F}{N}=\frac{E_{inf}}{N}=-T_c \log 2\;\;\;\;\;\;\;\; if \;\; T<T_c 
\nonumber \\
\nonumber \\
\frac{F}{N}=-T\log 2 \quad \quad \quad \quad \quad \quad \mbox{otherwise}
\end{array}\right.
\label{freemic}
\end{eqnarray}
This is a first order transition \cite{BouchMez}, unlike the standard
Random Energy Model transition, which is second order.
Let us stress here that above an  energy $E_{sup} =O(\log N)$
 the density of states is again much smaller than $1$. 
 At this level , independence of energies is no longer expected  \cite{Mertens}. 
In what follows, we shall be interested in transitions that
involve low energy states so that these levels are not relevant.
We shall not  study any statistical property of levels picked 
at random, this is done in Refs. \cite{Mertens,microsoft} and in all detail in  \cite{Silvio}.
We shall concentrate on the properties of energy levels {\em as encountered by a specific
dynamics}, a related though clearly inequivalent question.

\subsection{Dynamical analysis}

The equilibrium calculation shows  that the system maximises its entropy
at temperatures greater than $T_c$. 
In the low temperature phase $T \leq T_c$, a Metropolis dynamics of the system
explores deep states in its attempt to lower the energy down to $-\ln 2 \; N$.
We shall argue that single spin-flip  dynamics
naturally leads to a Trap Model, and perform numerical tests
to substantiate this claim.
The  Metropolis dynamics is defined by   
 transition rates verifying {\it detailed balance}:
\begin{eqnarray}
P(\{ \sigma_i \} \rightarrow \{ \sigma_j \}) = 
\left\{ \begin{array}{c}e^{- \beta (E_j-E_i)} \quad \mbox{if } E_j > E_i
 \\ 1  \quad \quad  \mbox{      otherwise} \\ \end{array}\right.
\label{Metr}
\end{eqnarray}
where $\{\sigma_i\}$ and $\{\sigma_j\}$ are configurations that differ by a single
spin flip. 
Below $T_c$ already an $N=30$ spin system cannot be equilibrated in reasonable computer time.

\subsubsection{Surface  states and horizon}

As mentioned above, one of the defining features of the trap model is a horizon level
to which the system has to return each time it escapes a trap.
Let us show that a single-flip dynamics naturally leads to this.
When 
\begin{equation}
E<E_h= T_c \log(a_{min}) \sim - T_c \log N 
\end{equation}
with $a_{\min} \equiv \min (a_1,...,a_N)=O(1/N)$, 
a
 single spin flip necessarily leads to a state whose energy is greater than $E_h$. 
In the present article, 
we call {\em trap} a state whose energy is lower than $E_h$ and {\em
surface state}, state whose energy is higher.
It will turn out that for long times the system is dominated by long stays in deep  traps,
 separated by rapid
excursions close to the horizon level.

\vspace{0.3cm}
\begin{figure}
\begin{center}
\psfig{file=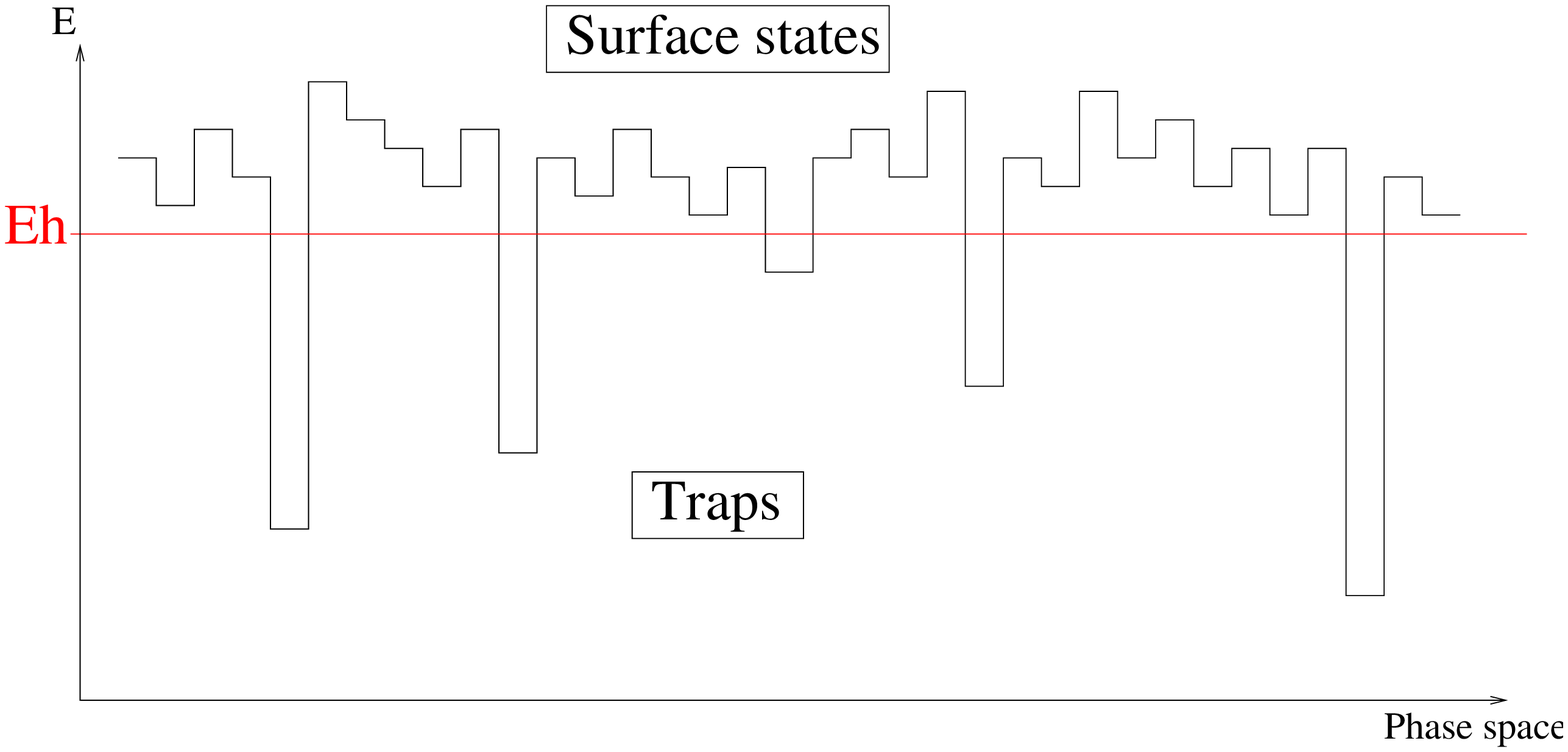,width=7.5cm,height=5.5cm}
\caption{Phase-space structure defining traps and surface states.}
\label{fig3}
\end{center}
\end{figure}

\begin{figure}
\begin{center}
\psfig{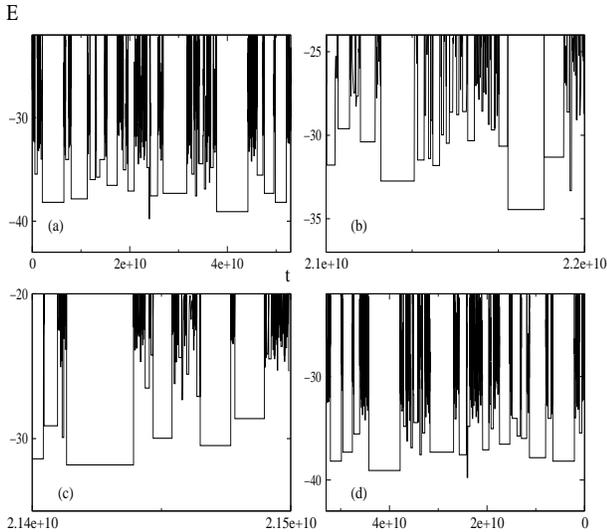}
\caption{Energy vs. time for a single run. (a)Whole run. (b)Time magnification by $10$
(c)Time magnification by $100$(d) inverse time}
\label{fig4}
\end{center}
\end{figure}

A quick first check can be obtained by looking at
 the 
energy evolution. Fig. \ref{fig4} shows a run for a $N=10000$ spins
system at temperature $T=0.75 \; T_c$. 
 It appears
that the expected scaling invariance of traps is well verified.
Also a necessary condition, the system looks statistically invariant with respect to 
 time reversal (cfr. \ref{fig4}(a) and \ref{fig4}(d)). Let us again stress this point:
the trap model is time-reversible 
in an interval between any two exits:
{\em once a trap is left the system is as unoptimized
as it was at the start}, when the system
is in a deep trap, it has no
other choice but to entirely reorganize itself to get to a new
deep trap.
An amusing exercise is to reconcile reversibility with the systematic trend
of energy decrease
we see in Figure \ref{fig5}.

\vspace{0.3cm}

The existence of the horizon level is a direct consequence of the one-flip dynamics
we have chosen, which forces deep traps to be many steps away from one another.
Indeed, the very same system allowed to flip $O(N)$ spins each step
has an entirely different dynamics: the jumps tend to go to  deeper levels at each
step, and Figures like  \ref{fig4} (a) and \ref{fig4} (d)
would look completely different.

\subsubsection{Time and energy distributions}

Let us first examine the transition rates in order to
determine the trapping time distribution.
In a first stage, we assume the system to be in a trap, that is $E < E_h$:
the only way to escape is to reach
a higher energy state. If we note  $x=\sum_{i=1}^N a_i s_i$,
the probability to escape from $x$ by a single spin flip reads:
\begin{equation}
P_{esc}(E(x))=\frac{1}{N}\sum_{i} 
e^{-\frac{T_c}{T}(\log\vert x-2a_i s_i \vert-\log\vert x \vert)}
\label{pesc1}
\end{equation}
If $\vert x \vert$  is much smaller than $a_{min}$ ($\iff E \ll E_h$):
\begin{equation}
P_{esc}(E)=\frac{A_N}{N}e^{\beta E}
\end{equation}
 $A_N \equiv \sum_i^N a_i^{-\frac{T_c}{T}}$ is the sum of $N$ random variables 
with distributions with power-law tails $p(X=a_i^{-\frac{T_c}{T}}) \sim X^{-(1+\frac{T}{T_c})}$ 
  for $T<T_c$. This implies   that for large $N$ it becomes distributed according
to a Levy law, and  hence \cite{BouGeo}  ${\bar{A}}_N \sim 
N^{\frac{T_c}{T}}$ and ${\bar{A}}_N^{-1} \sim 
N^{-\frac{T_c}{T}}$.
We thus have:
\begin{eqnarray}
\tau(E)&=&  \tau_0  e^{-\beta E} 
\label{tau}
\\
\tau_0 &\propto&   N e^{\beta E_h} 
\label{tau0} 
\end{eqnarray}
$\tau_0$ sets the timescale.
  Using the relation 
\begin{equation}
\psi(\tau) =  P(E) \vert \frac{\partial E}{\partial \tau} \vert
\label{relaproba}
\end{equation}
where $P(E)$ satysfies (\ref{energie}), we recover a Levy trapping 
time distribution:
\begin{equation}
\psi(\tau) \propto \frac{ \tau_0^{\frac{T}{T_c}}}{\tau^{1+\frac{T}{T_c}}}
\label{Levy}
\end{equation}

\begin{figure}
\begin{center}
\psfig{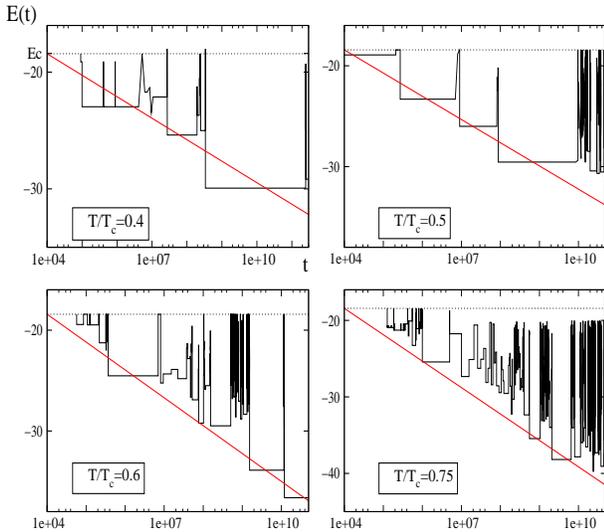}
\caption{Time-dependence of energy E(t) for single runs at different temperatures-$N=10000$. 
Straight lines are obtained from (\ref{Et})}
\label{fig5}
\end{center}
\end{figure}

If instead of assuming that one starts from a surface state, one computes the distributions
of energies (or, equivalently, trapping times $\tau$) as encountered at given time $t_w$, one
obtains distributions modified by the observation time.
If we consider the probability of being at $t_w$ in a trap near the horizon, it is 
reasonable to assume that these are populated with a probability proportional
to the Gibbs weight (with, of course, a cutoff at lower energies).
We hence have:
\begin{equation}
P_E(E,t_w) \propto e^{-(\beta-\beta_c)E} 
\end{equation}
which implies, using the change of variables $P_\tau(\tau,t_w)=P_E(E,t_w)\left| \frac{dE}{d\tau}\right|$:
\begin{equation}
P_\tau(\tau,t_w) \propto  \frac{\tau_o^{T/T_c-1}}{\tau^{T/T_c}} \;\;\; \;\; \tau \ll t_w
\end{equation}
In the opposite limit of deep energies, one may assume that two very deep states are visited
 with the same probability. One can prove this for the trap model, but here it remains a 
conjecture,
just like all independence properties. In any event, this would lead to:
\begin{equation}
P_E(E,t_w) \propto {\cal{N}}(t_w) e^{\beta_cE} 
\end{equation}
\begin{equation}
P_\tau(\tau,t_w) \propto \frac{ {\cal{N}}(t_w)}{\tau^{1+ T/T_c}} \;\;\; \;\; \tau \gg t_w
\end{equation}
demanding that $\int_{t_w}^\infty P_\tau(\tau,t_w) \; d\tau $ stays of order one,
 we have that $ {\cal{N}} \sim t_w^{T/T_c}$.
More generally, considering that surface states have also Gibbsean weights, 
one can write a formula valid in all regimes \cite{Bouchaud,BD}:
\begin{equation}
P_E(E,t_w) = e^{-(\beta-\beta_c)E} \; r\left(\frac{\tau_o e^{-\beta E}}{t_w}\right)
\end{equation}
where $r(u)=1$ when $u \ll 1$ and $r(u) \sim u^{-1}$ when $u \gg 1$.
In particular, we have on average:
\begin{equation}
\langle E(t)\rangle \sim -T_c \log \left(\frac{t}{\tau_0}\right)^{\frac{T}{T_c}}
\label{Et}
\end{equation}
where $\tau_0$ is given by (\ref{tau0}). 
In Fig. \ref{fig5}, we plotted different runs at different temperatures
for a system of $10000$ spins in order to show up the validity of (\ref{Et}). 
For different  $N$ similar curves are obtained.

\vspace{0.3cm}

\begin{figure}
\begin{center}
\psfig{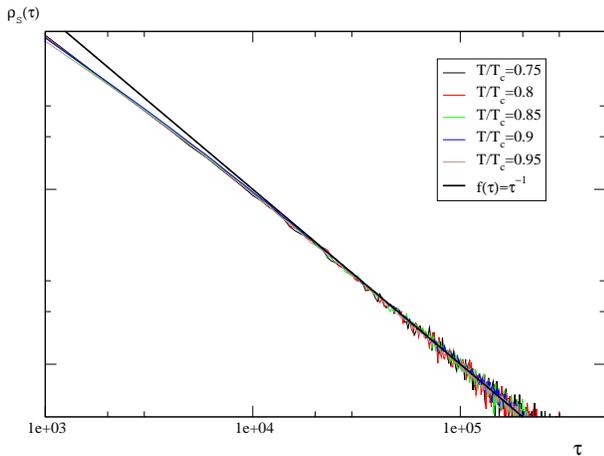}
\caption{Rescaled  trap time distribution-$N=1000$-$t_w=10^4$}
\label{fig6}
\end{center}
\end{figure}

\begin{figure}
\begin{center}
\psfig{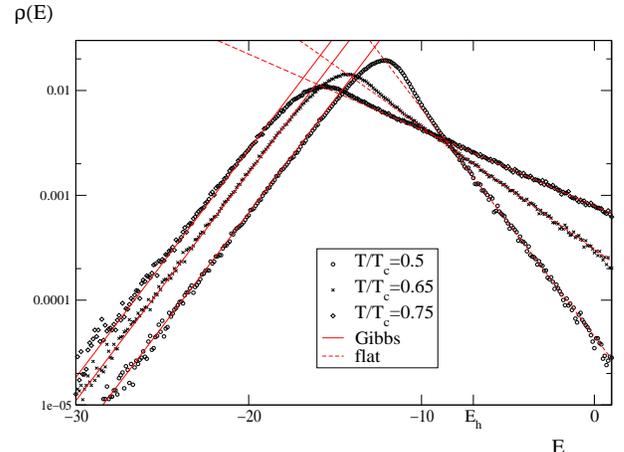}
\caption{Energy distribution for different runs of a $50$-spin system at $t_w=2 \; 10^4$.}
\label{fig7}
\end{center}
\end{figure}

At a given time, we computed the distribution of 
traps by counting the number $N_{\tau}$
of visited traps whose length is $\tau$. Fig. \ref{fig6} shows 
the rescaled density
$\rho_s(\tau)=\left(\rho(\tau)\right)^{\frac{1}{1+\frac{T}{Tc}}}$
for a system of $N=1000$ spins at $t_w=10^4 $ Monte Carlo steps,
 for different temperatures from $T=0.7 \; T_c$
to $T=0.95 \; T_c$. The curves are again in excellent agreement to what would have been obtained
for the trap model.
\vspace{0.3cm}
In Fig. \ref{fig7} we plot the energy of $50$ spins obtained at $t_w=2\; 10^4$ for different runs, at three
different temperatures. The straight lines correspond to the flat measure for the low-energy
tails, and the Gibbs distribution for the high-energy tails. The agreement with the trap model
behavior is excellent.
\vspace{0.6cm}

\subsubsection{Correlation between deep traps}

The correlation between states of given energy in the number partitioning problem
has been the object of detailed study following the proposal of Mertens 
\cite{Mertens,microsoft,Silvio}.
Although these studies are no doubt relevant to
the present work, one cannot transfer the results directly: while in a static study we 
consider the correlation of {\em any} two states below a certain  energy, we are here forced
to consider 
 the correlation between  states  as visited by the dynamics.
For example if we demand that two states are  visited subsequently, 
 given one state, we are considering a very specific subset for the next.
In other words, dynamics may impose correlations on statistically independent  energy levels, and
if independence emerges, it will be the result of a  property of the dynamics.

\begin{figure}
\begin{center}
\psfig{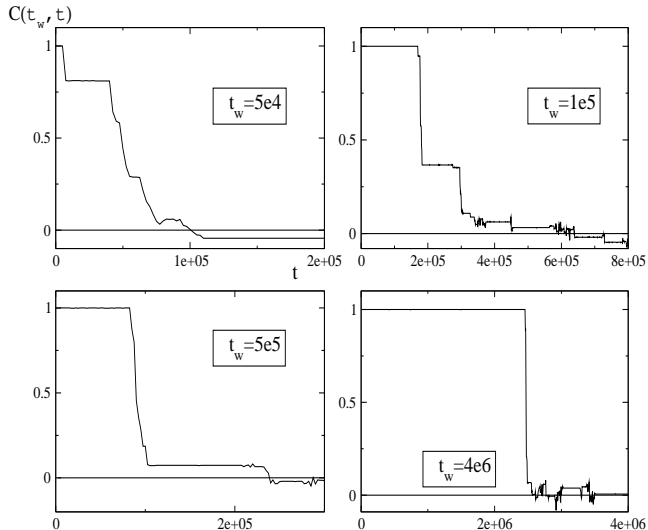}
\caption{Single run correlation functions $\frac{1}{N}\sum s_i(t+t_w) s_i(t_w)$ for
$N=1000$.$T=0.75T_c$. At longer times the bahvior approaches that of $C^B(t+t_w, t_w)$}
\label{fig8}
\end{center}
\end{figure}

Another, as we shall see related question is the following:
In the number partitioning model, the natural way to define the correlation
between configurations, based on the spins $s_i$ is: 
\begin{equation}
C_{Single}(t_w,t_w+t)=\frac{1}{N}\sum_{i=1}^{N} s_i(t_w) s_i(t_w+t)
\end{equation}
On the other hand, the usual correlation studied in the trap model is defined as:
\begin{equation} 
C_{Single}^B(t_w,t_w+t)= \left\{ 
\begin{array}{cc}
1 & \; \; if \;\; s_i(t+t_w)=s_i(t_w) \;\; \forall i \\
0  & \;\;\; otherwise
\end{array} \right.
\end{equation} 
We shall later consider also the average 
 correlation functions $C(t_w,t_w+t)=\langle C_{Single}(t_w,t_w+t) \rangle$
and $C^B(t_w,t_w+t)=\langle C_{Single}^B (t_w,t_w+t) \rangle$ 
($\langle \cdot \rangle$ denotes average over the noise history).
Due to the characteristics of the trap model, it turns out that in the low temperature phase, and 
for large waiting times, the two correlations coincide. The reason is interesting in itself:
at long times, the system spends most of the time in deep 
traps. Now even though on average  the passage between two deep traps  takes for larger
$t_w$ less and less 
proportion of the time, 
it still involves many spin flips.
Figure  \ref{fig8} shows this 
for different single runs with $N=1000$ at $T=0.75T_c$. 
 As $t_w \rightarrow \infty$, we see that the correlation $C_{Single}(t_w,t_w+t)$ becomes
essentially  a single jump process: this is because the route separating two traps of 
typical life $t_w$ becomes long.
\vspace{0.3cm}

\begin{figure}
\begin{center}
\psfig{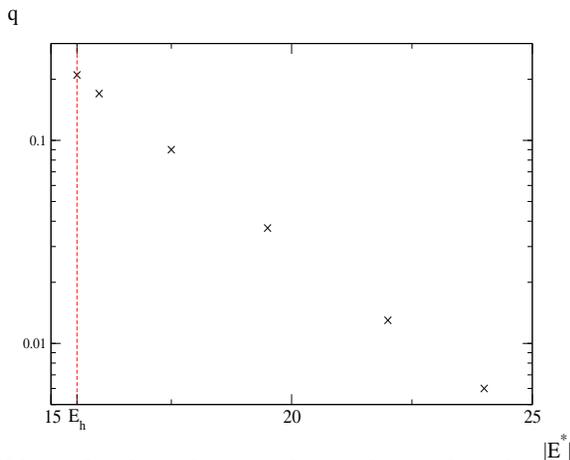}
\caption{Overlap $q$ between first and second configurations to be visited with $E<E^*$. 
$N=1000$.}
\label{fig9}
\end{center}
\end{figure}

Another way of confirming the increasing decorrelation 
during each passage over the horizon is to compute 
the first  ${\bf s}_{first}$ and second
${\bf s}_{second}$ configurations to be visited under the condition 
that their energies are smaller than a given energy
$E^*$. Their overlap
$q={\bf s}_{first} \bullet {\bf s}_{second}/N$ is the smaller, the lower the $E^*$ considered,
see Fig. \ref{fig9}.

\vspace{0.3cm}

We have also computed the averaged autocorrelation and the response function  
for a single sample:
\begin{eqnarray}
C(t_w,t_w+t)= \frac{1}{N}\sum_{i=1}^{N} \langle s_i(t_w)s_i(t_w+t) \rangle
\label{cor}
\\
R(t_w,t_w+t) \equiv \sum_{i=1}^N \frac{\partial \langle s_i(t) \rangle}{\partial h_i(t_w)}=
\int_{t_w}^{t_w+t} dt' \; \; 
\frac{\delta m(t)}{\delta h(t')}  
\label{rep}
\end{eqnarray}
The response is numerically obtained by computing $m(t)=\frac{1}{N}\sum_{i=1}^{N} \xi_i s_i(t)$ where
$\{\xi_i\}$ is a set of independent random variables that can take the
values $\pm 1$.
$h$ is an external field coupled to the spins via $\xi_i$. The
interaction term involved in Metropolis rates  reads:
\begin{equation} 
E_h=E+ V_h \;\;\; ; \;\;\; V_h= -h\sum_{i=1}^{N}\xi_i s_i
\label{VH}
\end{equation}

At high temperature ($T>T_c$), the functions (\ref{cor}) and (\ref{rep}) become
 time translational invariant: there is
no aging and the fluctuation dissipation theorem holds (see Fig. \ref{fig17}).

\begin{figure}
\begin{center}
\psfig{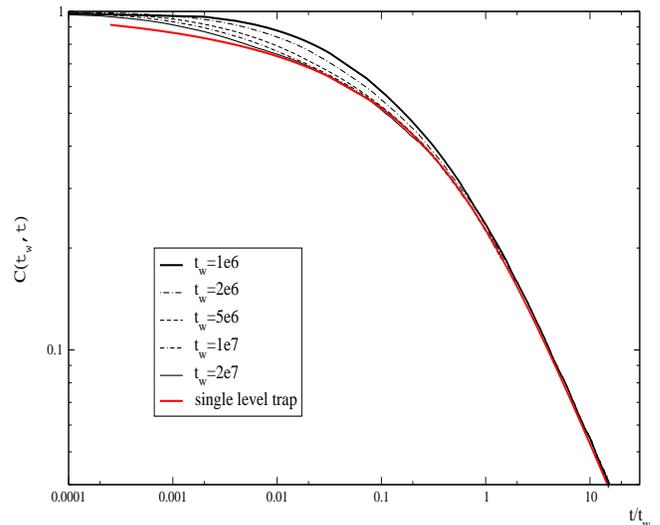}
\caption{Autocorrelation function-$N=1000$-$T=1.5$}
\label{fig10}
\end{center}
\end{figure}

At lower temperature, ($T<T_c$), the system is aging. We computed
the autocorrelation (\ref{cor}) for a $N=1000$ spins system at temperature
$T=0.75 \; T_c$ for different
waiting times $t_w$ as a function of $\frac{t}{t_w}$. The results are given
in Fig. \ref{fig10}. Firstly, we see that the longer we wait the better the scaling becomes.
We also see that the long-time behavior is well fitted
by the analytical results of Bouchaud {\it et al} \cite{BD,Cecile96}:
\begin{equation}
C(t_w+t,t_w) \simeq \frac{\sin \left( \pi \frac{T}{T_c} \right)}{\pi} 
\int_{\frac{t}{t_w+t}}^1 du \; (1-u)^{\frac{T}{T_c}-1}u^{-\frac{T}{T_c}}
\end{equation}
but there is a long preasymptotic subaging regime. 

\vspace{0.5cm}

In Fig. \ref{fig11},  we plot the autocorrelation (\ref{cor}) of
 $N=10000$ spins at $T=0.9 \; T_c$
for different waiting times $t_w$ as a function of 
($\frac{t}{t_w}$): the preasymptotic regime is much longer: $t_{sub} \sim  \; 10^7$ and
$t_{sub} \sim  \; 10^6$ for Figs. 
\ref{fig11} and Fig. \ref{fig10} respectively.

\vspace{0.6cm}
\begin{figure}
\begin{center}
\psfig{file=Correlation2.eps,width=7cm,height=8.5cm,angle=-90}
\caption{Autocorrelation fonction-$N=10000$-$T=1.8$}
\label{fig11}
\end{center}
\end{figure}
\begin{center}
\begin{figure}
\psfig{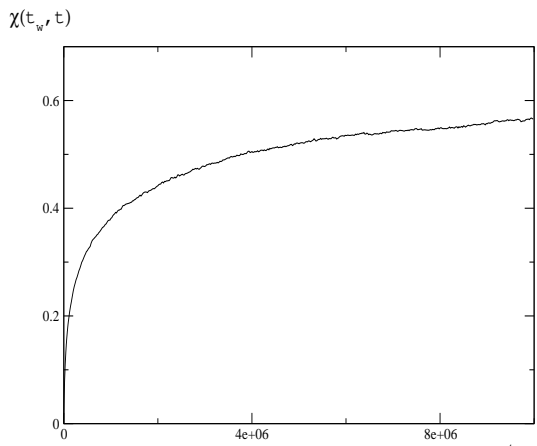}
\caption{Reponse function. $N=1000$. $T=0.75T_c$. $t_w=5e6$. $h=0.05$}
\label{fig12}
\end{figure}
\end{center}

Finally, in figure \ref{fig12} we also show a response function.

In this section, we have shown evidence  that  in the limit 
$t_w \rightarrow \infty$, the present  model becomes strictly the trap model:
short-time discrepancies  
are due to  the transition states in the frontier between
surface states and traps. Note however that near the critical temperature, critical corrections to
the asymptotical scaling $h(\frac{t}{t_w})$ are also expected in the original
Trap Model \cite{BPC}.

\subsubsection{Out of equilibrium fluctuation-dissipation relations}

One of the assumptions characterizing the trap model is the exponential
distribution of energies. Given its close ties to the mean-field picture, and more importantly,
to the `entropy crisis' scenario of glasses, this is a feature one is reluctant to give up.
On the contrary, when it comes to specifying how the transition times are
affected by a perturbation (such as a magnetic field) --- or equivalently, how
does a field affect the horizon level, there is considerable freedom.
Indeed, given two states  of energies $E$ and $E'$ with magnetizations  $M$ and $M'$, respectively, 
the transition probabilities can be chosen as:
\begin{equation}
P_{esc}(\{M,E\} \rightarrow \{M',E'\})=e^{-\beta h [-(1-\zeta)M'+\zeta M]} P_0(E \rightarrow E')
\end{equation}
where  
$P_0(E \rightarrow E')$ is the   rate  without external field
\cite{BD}. For any $\zeta$, detailed balance is obeyed.
One can thus show that the response becomes
\begin{eqnarray}
R(t_w,t_w+t)&=&\beta \left( -\zeta \frac{\partial C(t_w,t)}{\partial t} +
(1-\zeta) \frac{\partial C(t_w,t)}{\partial t_w} \right) \nonumber \\
&\sim& \beta \left[ \zeta \frac{t_w}{t} + (1-\zeta) \right]  
\frac{\partial C(t_w,t)}{\partial t_w}
\label{FDTb}
\end{eqnarray}
If $\zeta=0$  the rate is affected by the arrival configuration
(at first sight a bizarre choice), and the
 fluctuation dissipation formula
holds. If  $\zeta=1$, when the  rate depends on the departure configuration,
 there is a complicated fluctuation dissipation relation that
cannot be interpreted as resulting from an  effective 
temperature (see \cite{Sollich,Ritort} for a detailed discussion).

\begin{center}
\begin{figure}
\psfig{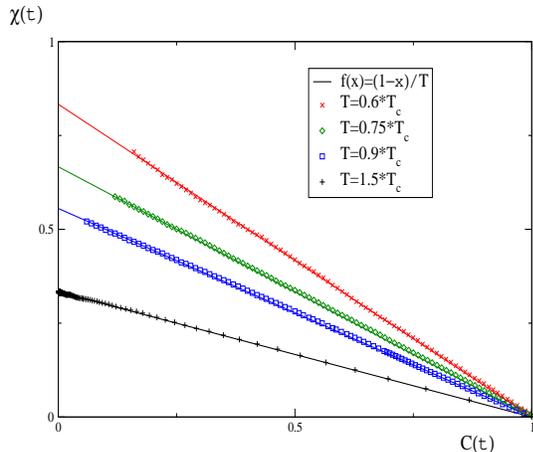}
\caption{Fluctuation-dissipation 
 relation for different temperatures above and below $T_c$. For $T=1.5T_c$, 
$N=10000$, $t_w=1e6$. For $T=0.9T_c$, $N=10000$, $t_w=5e6$. For $T=0.6T_c$,
$N=1000$, $t_w=5e4$.}
\label{fig17}
\end{figure}
\end{center}

The model studied here being `microscopic', we have no freedom to choose
how the magnetic field acts: the only reasonable choice is
that the rates are
given by the Metropolis prescription with an additional energy term
$V_h$ (\ref{VH}).
Fig. \ref{fig17} shows the fluctuation-dissipation characteristic at different 
temperatures above and below the critical temperature. Clearly,  
 there is no FDT violation~
\footnote{This result has already been found
in the one dimensional trap model \cite{BertinFDT}, see \cite{Monthus1D} for  
an analytic proof.}. 

As we shall see, one expects this result to be general for microscopic models
when the
  observables are  smooth functions of the spins.
To understand this  lack of violation,
first remark that it corresponds to the case $\zeta=0$, as if the transition time did not depend
on the original state. This is easy to understand: a smooth deformation of
a golf course does not change the depths of the holes with respect to their edge:
a single spin flip is enough to escape a  hole but it does  
 not change the magnetic energy! 

We have the following picture for state distribution:
 there are surface states and there are  traps that are reached from those 
states. Deep traps are separated by many surface (and short-lived) states, 
as we confirmed in the previous section. If the system is perturbed since the beginning 
by a small magnetic field,
there is a small reshuffling in energy, but surface states
remain on the surface.

Now, consider the time just before falling in a deep trap. The above
consideration  implies that the system is at the end of a tour
consisting of many short-lived surface states.
If we make the  natural assumption that such states follow
a Gibbs distribution even in the presence of a field, we conclude that
the magnetization distribution just before falling is:
\begin{equation}
P_{before}(M) = \frac{\int de \; P_{sup} (e,M) e^{\beta h M}}{Z_h} = \frac{ e^{\beta h M}
G(M)}{Z_h}
\end{equation}
where $ P_{sup} (e,M) $ is non zero only when $e > E_h$. $Z_h$
is the normalization factor.
After falling in the trap, we know that the energy has changed
dramatically. However, since the process of falling involves only one
spin flip, the
magnetization remains essentially unchanged (up to $O(1/N)$).
Hence,  the magnetization distribution inside the deep trap
is also of the form
\begin{equation}
P_{after}(M) =  \frac{ e^{\beta h M} G(M)}{Z_h}
\label{pafter}
\end{equation}

 From this, the FDT result follows, since:
\begin{center}
\begin{eqnarray}
\nonumber
  \langle M \rangle &=& \frac{\int dM \; M e^{\beta h M} G(M)}{Z_h}
\\ \nonumber
\langle M^2 \rangle - \langle M \rangle^2 &=&
\frac{ \int dM \; M^2 e^{\beta h M} G(M)}{Z_h} 
 \\
&-& \left( \frac{\int dM \;  M e^{\beta h M} G(M)}{Z_h}  \right)^2
\nonumber \\
&=& \beta^{-1} \frac{\partial \langle M \rangle}{\partial h}
\end{eqnarray}
\end{center}

We stress here that this argument will hold for {\em any} observable that,
unlike the energy itself, is {\em smooth} in phase space (i.e., 
such that  configurations that differ by a non-extensive number of spins 
have a  negligible difference in the observable -- in the opposite case,
we describe them as {\em rugged}). 

The fluctuation dissipation relation in our model comes from a Gibbsean
weight of states separating the traps. Numerical measurements
confirm this scenario:

\begin{enumerate}
\item
We computed the energy of the last state visited by the system 
before falling into a trap. Fig. \ref{fig18}
shows the results both with and without an external field.
We can see that there is
the announced zero field horizon value for 
$E_h=T_c \log(a_{min})$ under which the system cannot reach
a trap by a single flip. Comparison with   Fig. \ref{fig7}
 confirms that when an external field is added, 
the surface states essentially remain at the level of the zero-field surface, and
are consequently  still distributed following Gibbsean weights. 
Furthermore, the deviation with respect  to
$E_h$ vanishes faster than exponentially.

\begin{center}
\begin{figure}
\psfig{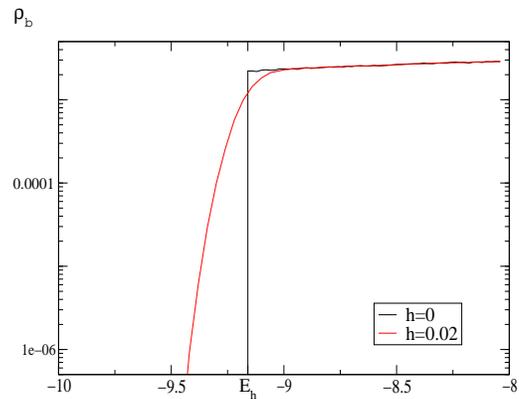}
\caption{Energy density of  states preceding traps. $N=50$. $T=0.75T_c$}
\label{fig18}
\end{figure}
\end{center}

\item
A direct consequence of our derivation is the
density probability (\ref{pafter}) for the magnetization. Fig. \ref{fig19}
confirms the scaling $P(M) \propto e^{\beta h M}$ for different values of $T$ and small
$h$.

\begin{center}
\begin{figure}
\psfig{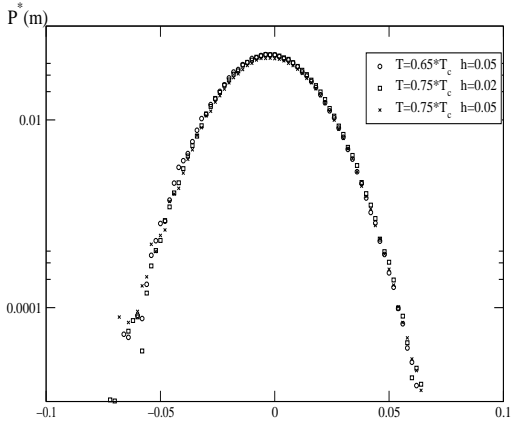}
\caption{Rescaled distribution $P^*(M)=e^{-\beta h M} P(M)$ for different values
of $h$ and $T$. Here, $m=\frac{M}{N}$ is the magnetization per spin. $N=50$}
\label{fig19}
\end{figure}
\end{center}

\end{enumerate}

\begin{center}
\begin{figure}
\psfig{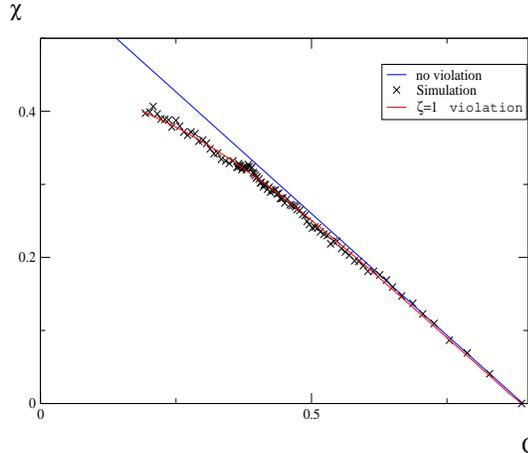}
\caption{fluctuation-dissipation 
relation for observable (\ref{MT}). $N=50$. $T=1.5$. $t_w=5e4$. $h=0.005$}
\label{fig20}
\end{figure}
\end{center}

From our analysis, the fluctuation-dissipation relation holds for smooth observables. The
situation is different in the case of rugged observables.
To verify this, we computed the fluctuation-dissipation 
relation in the case of an observable 
 that resembles 
the magnetization but is restricted to be non zero only when the
energy configuration is less than $E_h=-T_c \log(a_{min})$:
\begin{equation}
{\tilde M}(\{ s_i \})=
\left\{ \begin{array}{c} \sum_{i=1}^N \xi_i s_i \quad \mbox{ if  } E(\{ s_i \}) < E_h
 \\ 0  \quad \quad  \mbox{      otherwise} \\ \end{array}\right.
\label{MT}
\end{equation}
The fluctuation-dissipation 
plot is shown Fig. \ref{fig20}. The violation follows the relation:
\begin{equation}
{\tilde R}(t_w,t)=-\beta \frac{\partial {\tilde C}(t_w,t)}{\partial t}
\label{FDT1}
\end{equation}
which corresponds to the case of $\zeta = 1$.
This result is not surprising: On one hand, since ${\tilde M}$ vanishes
in the surface states it does not affect the dynamics between traps
 and is insensitive to the depth of
 the arrival trap. On the other hand, the probability to
escape from a trap with  ${\tilde M}$ is given by:
\begin{equation}
P_{esc}({\tilde M},E) \propto e^{\beta (E-h{\tilde M})} 
\end{equation}
so that we have $\zeta=1$. 

\begin{center}
\begin{figure}
\psfig{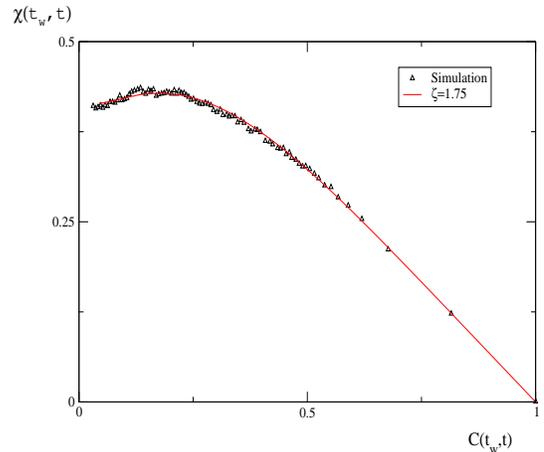}
\caption{FD relation in the case of Parity observable}
\label{fig21}
\end{figure}
\end{center}

We are now able to give a physical picture  of the fluctuation-dissipation relations
for all $\zeta$. Consider an observable $A$ that is the combination of two observables
  $A_1$ rugged (as  (\ref{MT})) and   $A_2$ smooth:
\begin{equation}
A=\zeta A_1 +(1-\zeta) A_2
\end{equation}
and $\langle A_1^2(t_w) \rangle = \langle A_2^2(t_w) \rangle \rightarrow 1$.
By definition
$A_1$ verifies (\ref{FDT1}) and the equilibrium fluctuation-dissipation relation holds for
$A_2$. 
Using linearity  and the fact that for long times the autocorrelations
of $A_1$ and $A_2$
respectively become the same  (a consequence of the dynamics 
of the trap model) one easily recovers the general case (\ref{FDTb}).
Let us finally 
 point out that the fluctuation-dissipation relation (\ref{FDTb}) is not restricted
to values of $\zeta$ between $0$ and $1$. 
 As an example, we report in Fig. \ref{fig21} 
the fluctuation-dissipation relation for the parity: 
\begin{equation}
P(\{ s_i \})= (-1)^{\frac{1}{2}\vert \sum_{i=1}^{N} s_i \vert} 
\end{equation}
It yields fluctuation-dissipation plots  with (\ref{FDTb})  $\zeta \sim 1.75$. 

\subsubsection{Nonexponential decay above $T_c$}

Trap models exhibit interesting equilibrium dynamics in a range above $T_c$.
Let us see how this comes about in our microscopic model. Consider
first the canonical partition function associated with (\ref{Ham}):
\begin{equation}
Z \sim \int_{E_{inf}}^0 dE  \; \; e^{N \log 2 -(\beta -\beta_c) E}
\end{equation}
where $E_{inf}=-NT_c\log2$.
We make the  change of variables:
\begin{equation}
n(E)=e^{-\beta(E_h-E)}
\end{equation}
$n(E)$  represents the probability
of escaping from a trap whose energy is $E$ 
with  the local dynamics (\ref{Metr}).
Then, $Z$ can be rewritten as:
\begin{equation}
Z \sim e^{N \log 2-(\beta-\beta_c)E_h} \int_{e^{-\frac{T_c}{T}N \log2}}^{e^{-\beta E_h}}
dn \;\; n^{\frac{T}{T_c}-2}
\end{equation}
Since $E_h$ is not extensive and is independent of $T$,
the free energy in the
thermodynamic limit can be written as:
\begin{eqnarray}
\nonumber
F&=&D-TN\log2 \\
\nonumber
D&\equiv&-T\log \left( \int_{2^{-\frac{T_c}{T}N}}^1 
dn \; f(n) \right) \\
f(n)&=&n^{\frac{T}{T_c}-2}
\end{eqnarray}
Three cases must be considered:
\begin{itemize}
\item{{\bf $T>2T_c$}\\
In this case $\int_0^1 dn \; f(n)$ is finite and  $D$
has a non-extensive contribution to the free energy. 
}
\item{{\bf $Tc<T<2T_c$}\\
$\int_0^1 dn \; f(n)$ is still finite and $D$ has a non-extensive
contribution, but $f(n)$ has a singularity at $n=0$:
\begin{figure}
\begin{center}
\psfig{file=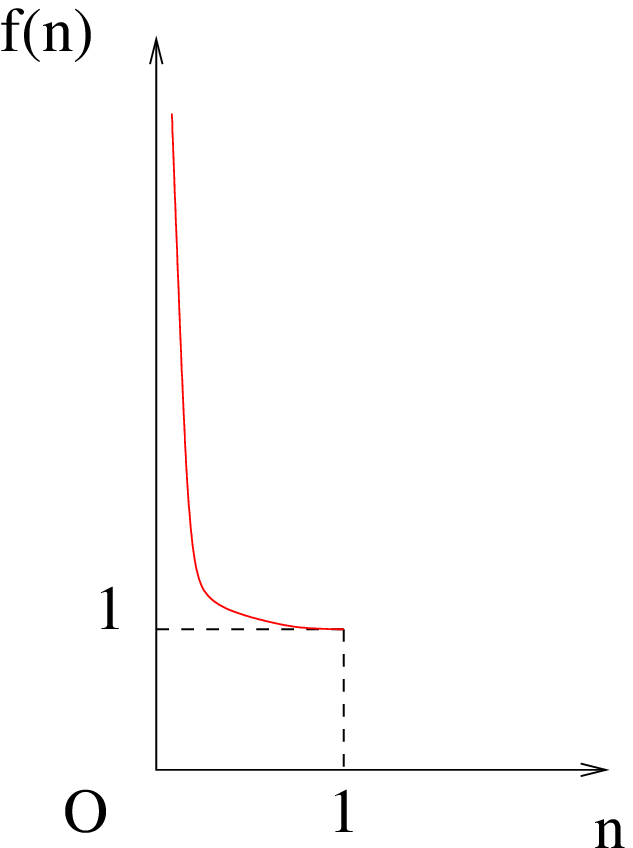,width=4cm,height=4cm}
\label{fig22}
\end{center}
\end{figure}
}
\item{{\bf $T<T_c$}\\
This time $\int_{\epsilon}^1 dn \; f(n)$ diverges when $\epsilon \rightarrow 0$,
and hence  $D$ has an extensive contribution to the free energy.
}
\end{itemize}
We see that below $2T_c$ the equilibrium measure   abnormaly
populates the dynamical states with  low probability of escape
$n(E) \sim 0$.  This is the origin of time-heterogeneites
in dynamics and it naturally coincides with the divergence
of trapping times'  variance (\ref{Levy}). As $T$ is lowered, 
the effect of low $n$ states  in the dynamics becomes more and more
pronounced \cite{Odagaki,Cecile96,Reichmanbouchaud,Ludovic},
and the equilibration time finally diverges at $T_c$,
when the states with low $n(E)$ become dominant.
In Fig. \ref{fig23} we compare the correlation function at $T=\frac{3}{2} T_c$ and
at $T=2 T_c$. Although there is no
aging and the system equillibrates, the correlation function has a long tail in the former,
that is absent in the latter case.

\begin{figure}
\begin{center}
\psfig{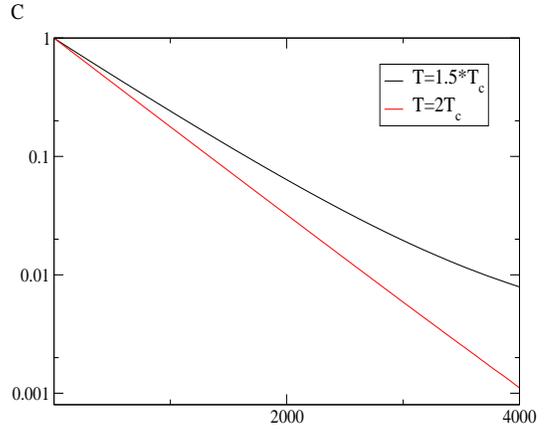}
\caption{Correlation function above $T_c$.$N=1000$}
\label{fig23}
\end{center}
\end{figure}

\subsubsection{Other trap distributions}

The Number Partitioning model can also be used as a microscopic basis for modified trap models.
One may  consider the  definitions for the energy:
\begin{equation}
E^{\hat \zeta}= \left| \sum_i a_i s_i \right|^{-(1+{\hat \zeta})} 
\end{equation}
for ${\hat \zeta}>0$, leading to
non-extensive cost functions.
 Metropolis dynamics with this energy still goes in the right direction,
but the dynamics is {\em not} trap-model like. In fact, 
repeating the arguments above one finds trapping times  distributed
according to $P(\tau) \propto \tau^{-1} [\ln(\tau/\tau_o)]^{(1+{\hat \zeta})}$.
This case has also been discussed by Bouchaud \cite{Bouchaud}: at any temperature, 
one observes traps that become 
systematically longer in time.

\section{Diophantine approximations}
  
\begin{figure}
\begin{center}
\psfig{file=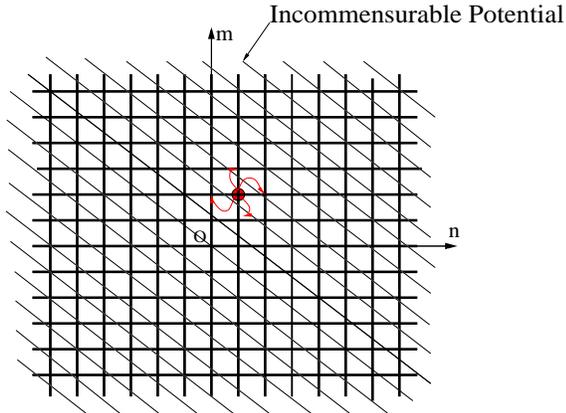,width=7.5cm,height=5.5cm}
\caption{2d-diffusion with irrational steps. The diagonal lines represent
the integer values of $n \sqrt{2}+m \sqrt 3 = \; even$, where the energy is $-\infty$.  }
\label{fig24}
\end{center}
\end{figure}

\begin{figure}
\psfig{file=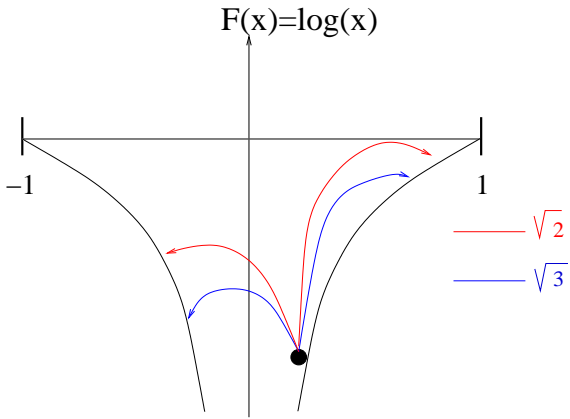,width=7.5cm,height=5.5cm}
\caption{1d-diffusion with irrational steps.}
\label{fig25}
\end{figure}

Number theory is a gold mine for glassy models without quenched randomness.
This is because a  program failing to find the good solution is trapped by 
the usually enormous number of near misses which behave as quasi-random
numbers.
Consider the problem of Diophantine approximations: we are asked, for example, to
find integers $n$ and $m$ so that $n \sqrt{2}+m \sqrt 3$ is as close as possible to an
even integer. We can express this by saying that we want to minimize an energy
\begin{equation}
E=\ln \{ F(n \sqrt{2}+m \sqrt 3+\frac{1}{2})\}
\label{dos}
\end{equation}
where $F(x)$ is the separation between $x$ and its nearest even integer.
One can
  view this optimization problem as a
diffusion
of a particle in a bidimensional lattice $(n,m)$
with a periodic potential (\ref{dos}) inconmensurable with the lattice spacing 
(Fig. \ref{fig24}), so that the places with $E=-\infty$ are always missed.
\vspace{0.3cm}
Let us give an alternative representation.
Fig. \ref{fig25} shows the same problem, where  we have now plotted one period
of the potential, and assumed periodic boundary conditions.
\vspace{0.3cm}
 In this representation, {\em each move is a large step}, because
it goes around the boundary.
What this second representation suggests is that a succession of many moves
amounts to generating a (pseudo) random number in the interval $(-1,1)$:
we can thus expect the equilibrium measure to be flat and $E=\ln |x|$.
The energy density of states of energy $E$ would then be exponential, since:
\begin{equation}
P(E)  =  \; \left| \frac{dx}{dE}\right| \sim e^{E}
\end{equation}
Now, given a point  $(n,m)$ that yields a good approximation to an even integer,
moving to nearby points  like $(n+1,m)$ or $(n,m+1)$ will completely spoil
the approximation, since it implies jumps of $\sqrt 2-1$ or  $\sqrt 3-1$.
{\em This is an horizon as in the previous section: it is necessary to reemerge
in order to find a deeper place}.~\footnote{The situation is strongly reminiscent
of the trapping in   subrecoil laser cooling,
see \cite{laser}}.

\begin{figure}
\begin{center}
\psfig{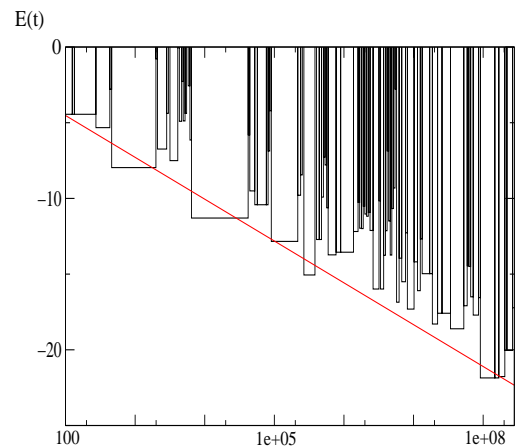}
\caption{Approximating integers as $p\sqrt{2}+q\sqrt{3}+r\sqrt{5}$, with $p,q,r$ integers
 (see text) }
\label{fig26}
\end{center}
\end{figure}

We can now make the assumption that in the large, the diffusion process as viewed in
Fig. \ref{fig24} is a diffusion in a lattice with trapping times distributed according to a 
L\'evy law.  From what we know from such problems, returns are relatively
frequent in two dimensions, somewhat changing the behavior \cite{Cecilebasse}.
In order to make the comparison simpler, we have thus simulated a three dimensional problem
$E=\ln F(p \sqrt 2 + q \sqrt 3 + r \sqrt 5)$. Figure \ref{fig26} shows the 
behaviour of energy which looks just like in a trap model, the
 Levy flight leading to a subdiffusion process \cite{Cecilebasse}.
 We have performed most
of  the tests as in the previous section, but the results being numerically indistinguishable,
we do not present them here.  
\vspace{0.3cm}

In order to stress the relevance of the horizon in the dynamics, consider
the potential $\ln |x|$ with $x \in (-1,1)$, with Metropolis dynamics but with
the configurations drawn each time at random in the interval  $(-1,1)$.
One can easily check that the dynamics is {\em entirely different} from the trap model.
The best way to convince oneself of this is to consider the limit $T=0$: unlike the trap
model, the system manages to decrease its energy monotonically without any activation,
it suffices to wait long enough that the new configuration proposed has
a sufficiently low new energy.
 On one hand, we have
a Bouchaud dynamics for the irrational jumps and on the other hand, we have
a diffusion that resembles more to the Barrat Mezard model (where there is only descent)
 \cite{BM} if all steps are allowed. 
In the case of very small irrational  steps, we can expect a crossover between
these two regimes \cite{BertinCROSS}.

\pagebreak

\section{Discussion}

\setcounter{subsubsection}{0}

\subsubsection{Trap model behaviour}

In this paper we have given two instances in which trap model like behaviour arises
at long times, and we have used them to  draw conclusions on
what are the features required from a microscopic model for this to happen.
A first obvious condition is the existence of states with a large distribution of trapping
times. This, however, is not enough.
As mentioned several times in this paper, the trap model is such that once
 system emerges from a trap to a `horizon' level,
 it is completely  reinitialized. Furthermore, it is reversible in the 
sense that given a time interval delimited by two escapes, the history within it is equally 
probable than its time-reversed one.
 Irreversibility only arises because 
the fraction of time the system spends near the horizon becomes progressively smaller,
although it never vanishes.

The  question one may ask is under what circumstances can one have a horizon level with 
such a property. Considering the Number Partitioning problem
as the problem of minimising the height of a box needed
to pack two piles of coins of random thicknesses, we have seen that the horizon level arises 
when the system has been optimized up to the thickness of the thinnest coin: after that,
any swap of coins
will necessarily bring back the system to the horizon, and improvements only result after
global rearrangements.
One has then a crossover between 
irreversible early dynamics, where single swaps may be advantageous, to a trap/reinitialization
dynamics at longer times.
One may conjecture that this might be quite general of packing problems.

\vspace{0.3cm}
\begin{figure}
\begin{center}
\psfig{file=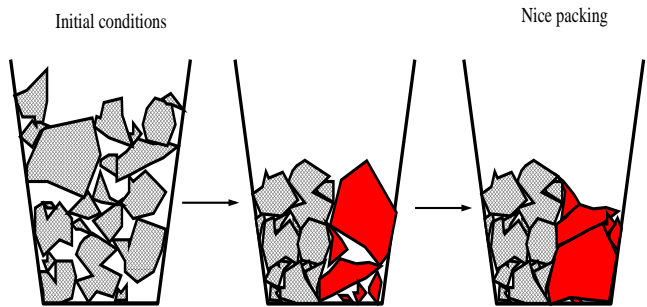,width=8.5cm,height=4.cm}
\caption{Direct compaction (left) at the first stages, global rearrangements
as in the trap behaviour at late stages (right).}
\label{fig27}
\end{center}
\end{figure}

Glass systems and granular matter have no doubt a first regime of irreversible compaction, 
during  which a trap (whenever we are able to define one) is most probably followed
by a deeper trap, and this is also the case of the dynamics of schematic  models.
Thus, one can envision true Trap Model behaviour arising at extremely long times,
when each improvement requires complete reshuffling: this will surely happen at
the level of optimization of a few grains (Fig. \ref{fig27}), in agreement
with the analysis for  the Random Energy Model \cite{Benarous}.
 Another, perhaps more relevant situation could 
arise at shorter times, but considering the trap behaviour of subsystems separately.

\subsubsection{Fluctuation-dissipation relation}

One of the main conclusions  of the analytic solutions of glassy dynamics \cite{review}
is the central role played by   the fluctuation-dissipation relations. It has hence become
standard practice to study numerical and experimental systems from this point of view, and
quite naturally one is led to look at this question in the Trap Model. 
As we have mentioned above, the  model as it stands allows for great freedom
in this respect: since we are free to specify how the fields modify the barriers,
the response to a field can strongly depend on this prescription.
One of the main points of this paper is that for observables that are smooth functions of the 
microscopic variables, the 
fluctuation-dissipation relation holds in the aging phase just
as in equilibrium. Although we have shown this for the two specific models, the argument
seems robust enough: the energy being rugged, escaping a deep trap involves
a few steps, a distance along which a smooth variable does not vary. Only long trajectories feel
spatially smooth perturbations, and these happen near the horizon level which can be assumed to be
in local equillibrium.
Let us note here that 
smooth variables cannot be correlated with the energy, which is itself rugged, so they
are `neutral' in the sense of Sollich \cite{Sollich}: they are not `what is being compactified'.

Once again we find confirmation that trap model behaviour can only arise at very long times: 
for example in mean-field models even within the activated regime one has initially
a violation of the equilibrium fluctuation-dissipation relation \cite{Crisantiritort},
and the emergence of an effective temperature.
This means that if these models eventually cross over to trap model behaviour, this will be
only after the effective temperature has thermalised with the bath temperature, and this
is expected to happen when energies are barely (to $O(N)$) above the equilibrium energy.
It would be interesting to understand this crossover better, as it may be relevant for 
finite-dimensional models.  

\subsubsection{Optimization and Non-extensivity.}

The Number Partitioning problem is nonextensive if one defines the energy
as the absolute value of the difference, in the sense that the ground state
energy scales as an exponential of the size, or equivalently, that in a
thermodynamic construction the interesting temperatures are exponentially close to zero.
Working with temperatures that depend on the size is always awkward, so we have chosen 
a new energy as the logarithm of the old one.
This immediately led to a well defined thermodynamics and, via Metropolis dynamics,
to trap model behaviour.

Because the trap model is by construction forgetful of its history, its appearence
in an optimization algorithm is a sign that things are as bad as possible.
Indeed, our scaling forms for the dynamics  immediately yield exponential
times to reach the ground  state - essentially what one would have obtained by 
blind enumeration.
An interesting property one can check is that, not surprisingly
{\em the best temperature from the point of view of optimization is the critical temperature}.


\vspace{0.5cm}
{\bf Acknowledgements}\\
We wish to thank E. Bertin and J-P. Bouchaud for useful suggestions, and
S. Franz for discussing his unpublished results.

\end{document}